\documentstyle[prl,twocolumn,aps,epsf]{revtex}
\begin{document}
\twocolumn[\hsize\textwidth\columnwidth\hsize\csname@twocolumnfalse%
\endcsname
\title{Effects of Fermi energy, dot size and leads width on weak localization \\
in chaotic quantum dots}

\author{E. Louis}

\address{Departamento de F{\'\i}sica Aplicada,
Universidad de Alicante, Apartado 99, E-03080 Alicante, Spain.}

\author{J.A. Verg\'es}

\address{{I}nstituto de Ciencia de Materiales de Madrid,
Consejo Superior de Investigaciones Cient\'{\i}ficas,
Cantoblanco, E-28049 Madrid, Spain.}

\date{\today}

\maketitle

\begin{abstract}
Magnetotransport in chaotic quantum dots at low magnetic fields is investigated
by means of a tight binding Hamiltonian on $L \times L$ clusters
of the square lattice. Chaoticity is induced by introducing 
$L$ bulk vacancies. The dependence of weak localization on the Fermi
energy, dot size and leads width is investigated in detail and
the results compared with those of previous analyses, in particular
with random matrix theory predictions. Our results indicate that 
the dependence of the critical flux $\Phi_c$ on  the square root of 
the number of open modes, as predicted by random matrix theory, is 
obscured by the strong energy dependence
of the proportionality constant. Instead, the size
dependence of the critical flux predicted by Efetov and random matrix theory, 
namely, $\Phi_c \propto \sqrt{1/L}$, is clearly illustrated by 
the present results. Our numerical results do also show that the weak 
localization
term significantly decreases as the leads width $W$ approaches $L$. However,
calculations for $W=L$ indicate that the weak localization effect
does not disappear  as  $L$ increases.
\end{abstract}

]

\narrowtext
\section{Introduction}
Experimental studies of magnetoconductance in quantum dots show
that, at low magnetic fields (typically below one flux quantum), the 
conductance increases with the field 
\cite{MR92,CB94,Ta97,SK98}. 
The effect has been investigated theoretically \cite{CB94,BJ93,YI95,GM98} 
and related to a similar
behavior observed in disordered metallic conductors in the diffusive
regime, that is referred to as Weak Localization (WL)
\cite{LR85,Ef97}.

There is also fairly conclusive experimental evidence which indicates that 
the average magnetoconductance $<G(B)>$ behaves in a qualitatively
different way in regular and chaotic cavities, namely, whereas
in the former it increases linearly with $B$, in chaotic cavities 
the WL peak has a Lorentzian shape \cite{CB94}. 
Semiclassical analyses ascribe this difference to the distributions 
of the areas $A$ enclosed by
the trajectories of the carriers \cite{BJ93}. While in
regular systems the probability distribution of enclosed areas larger than
$A$ is $\propto 1/A$ \cite{LB92}, in fully chaotic systems it is 
exponential \cite{JB90}.  
As a consequence, in chaotic cavities the increment in the 
magnetoconductance  as a function of magnetic flux $\Phi$ is given by:
\begin{equation}
\delta G=G(\Phi)-G(0) = \frac{a \Phi^2}{1+b\Phi^2} \;,
\label{e:peak}
\end{equation}
\noindent where the conductance and the magnetic flux are given in units
of their respective quanta, $G_0=e^2/h$ and $\Phi_0=h/e$.
The constant $b$ gives the critical flux at which the time--reversal
symmetry is effectively destroyed, $\Phi_c =1/\sqrt{b}$, whereas the ratio
$a/b$ gives the weak localization term, i.e., $G(\infty)-G(0)=a/b$.
The supersymmetric $\sigma$--model  predicts that 
$a$ and $b$ should be inversely proportional to the number of channels 
$N_{ch}$ that contribute to the current \cite{Ef97},
\begin{equation}
b = c\frac{\nu D_0}{N_{ch}} \propto \frac{L}{N_{ch}}
\label{e:b}
\end{equation}
\noindent where $c$ is a constant (for the present geometry $c=2\pi/3$), 
$\nu$ the density of states, $D_0$
the diffusion coefficient and $L$ the linear size of the cavity.
The size dependence arises  from the standard expression for the diffusion 
coefficient $D_0 =v_F l/2$, where
$v_F$ is the Fermi velocity, and $l$ the elastic mean free path, and 
from the fact that
in a two--dimensional ballistic system, $l \propto L$ \cite{LC97}. 
The qualitative behavior of Eq.~(\ref{e:b}) is similar to the Random 
Matrix Theory (RMT) result  for the critical
flux at which the time--reversal symmetry is broken (GOE--GUE transition)
reported in \cite{Be97}. 
The two constants $a$ and $b$ are proportional to each other. In
particular RMT gives \cite{GM98,PW94},
\begin{equation}
\frac{a}{b} = \frac{N_{ch}}{4N_{ch}+2}\;,
\label{e:RMTab}
\end{equation}
\noindent where $N_{ch}$ is related to the zero field conductance through,
\begin{equation}
G_{\rm RMT}(0) = \frac{N_{\rm ch}}{2}-\frac{N_{\rm ch}}{4N_{\rm ch}+2},
\label{e:RMTnch}
\end{equation}
\noindent On the other hand, a fitting of the numerical results obtained from
a random matrix model Hamiltonian gave \cite{PW94}
\begin{equation}
b = 2k\frac{2N_{ch}-1}{N_{ch}^2}\;,
\label{e:RMTb}
\end{equation}
\noindent   
$k$ being a constant which, as in Eq.~(\ref{e:b}), depends on the 
Fermi energy. Although Eq.~(\ref{e:RMTb}) gives the same dependence on 
the number of channels than Eq.~(\ref{e:b}) in the large $N_{ch}$ limit, 
it does not explicitly reproduce neither its size nor its energy dependence. 
Moreover, as remarked in \cite{PW94}, Eq.~(\ref{e:RMTb}) is only valid 
for few channel
ballistic cavities. It should also be mentioned that the size dependence of
the constant $b$  has also been obtained within RMT (see \cite{FP95,PW95}).

At present there is no published numerical study of the effects of the size of 
the cavity, the leads width, and the Fermi energy
on weak localization in reasonably realistic models of quantum chaotic 
cavities. The purpose of this work is to discuss the results of such an 
investigation. Quantum dots are described by means of a tight-binding 
Hamiltonian
on $L \times L$ clusters of the square lattice. Non--regular (chaotic)
behavior is induced by introducing  a number of bulk vacancies
proportional to the linear size of the system \cite{VL99}.
This model has been shown to behave similarly  to dots
in which chaoticity is induced by introducing disorder at the surface
\cite{CL96,BM98}.  The effects of leads width $W$, system size, and 
number of channels that contribute to the current 
are discussed in detail. Our results show that 
the the critical flux is not simply proportional to the square
root of the number of open channels as concluded in \cite{PW94}; it turns
out that this
relationship is obscured by the strong energy dependence of the 
proportionality constant already implicit in Eq.~(\ref{e:b}). 
Significant deviations from RMT are observed for large leads width 
($W$ of the order of the system size $L$). In particular the weak localization 
term decreases as $W$ approaches $L$. However, our numerical data 
for $W=L$ indicate that this term does not vanish as $L$ increases.

The paper is organized as follows. Section II includes a description
of our model of chaotic quantum dot and of the method we used
to compute the current. The results are discussed in Section III. We
first briefly consider the case of zero field, comparing our results
with those derived from RMT. The results concerning the effects
of Fermi energy, leads width and dot size are presented and discussed
thereafter. Again, comparison with RMT is highlighted. Section IV is devoted
to summarize the conclusions of our work.

\section{Model and Procedures}
\subsection{Model of quantum chaotic dot}
Our model of a quantum chaotic dot is described by means of a tight--binding 
Hamiltonian with a single atomic orbital per lattice site,

\begin{eqnarray}
\widehat H= - \sum_{<m,n;m',n'>} t_{m,n;m',n'}|m,n><m',n'|,
\end{eqnarray}

\noindent
where $|m,n>$ represents an atomic orbital on site $(m,n)$.
Indexes run from 1 to $L$, and the symbol $<>$ denotes that the sum
is restricted to the {\it existing} nearest-neighbors of site $(m,n)$. 
Using Landau's gauge the hopping integral is 
$t_{m,n;m',n'}={\rm exp}\left (2\pi i \frac{m}{(L-1)^2}\frac{\Phi}{\Phi_0}
\right )$, for $m=m'$, and 1  otherwise. 
Therefore, the difference between our Hamiltonian $H$ and the one 
corresponding to an ideal $L \times L$ cluster on the square lattice is the
absence of hopping to and from $L$ sites chosen at random among the $L^2$
sites defining the lattice. A full discussion of the properties of this
model for the case of a closed system and  zero field can be found
in Ref. \cite{VL99}.

\subsection{Conductance}
The conductance (measured in units of the quantum of conductance
$G_0 = e^2/h$) was computed by using the  implementation
of Kubo formula described in Ref. \cite{Ve99} (applications to mesoscopic
systems can be found in Refs. \cite{CL97} and \cite{LV00}). 
For a current propagating in the $x$--direction,
the static electrical conductivity is given by:

\begin{equation}
G = -2 {{\left( \frac{e^2}{h} \right)} {{\rm Tr} \left [(\hbar {\hat v_x})
{\rm Im\,}{\mathcal \widehat G}(E)(\hbar {\hat v_x})
{\rm Im\,}{\mathcal \widehat G}(E)\right ]}} \;,
\end{equation}
where ${\rm Im\,}{\mathcal \widehat G}(E)$ is obtained from
the advanced and retarded Green functions:
\begin{equation}
{\rm Im\,}{\mathcal \widehat G}(E)=\frac{1}{2i}\left[{\mathcal
\widehat G}^{R}(E)-{\mathcal \widehat G}^{A}(E)\right ] \;,
\label{e:img}
\end{equation}
and the velocity (current) operator ${\hat v_x}$ is related
to the position operator ${\hat x}$ through the equation of motion
$\hbar {\hat v}_x = \left [ {\widehat H},{\hat x} \right ]$,
$\widehat{H}$ being the Hamiltonian.

Numerical calculations were carried out connecting quantum dots
to semiinfinite leads of width W in the range 1--$L$.
The hopping integral inside the leads and between leads and dot
at the contact sites is taken equal to that in the
quantum dot (ballistic case). Assuming the validity of both
the one-electron approximation and linear response,
the exact form of the electric field does not change the value of $G$.
An abrupt
potential drop at one of the two junctions provides the simplest
numerical implementation of the Kubo formula \cite{Ve99} since, in this case,
the velocity operator has finite matrix elements on only two adjacent
layers and Green functions are just needed for this restricted
subset of sites. Assuming this potential drop to occur at the left contact
($lc$) side, the velocity operator can be explicitly written as,
\begin{equation}
i\hbar v_x = -\sum_{j=1}^W\left (|lc,j><1,j|-|1,j><lc,j|\right)
\end{equation}
\noindent where $(|lc,j>$ are the atomic orbitals at the left contact sites
nearest neighbors to the dot.

Green functions are given by:
\begin{equation}
[E \widehat I - \widehat H - \widehat \Sigma_{\mathrm 1}(E) - \widehat
\Sigma_{\mathrm 2}(E)] {\mathcal \widehat G}(E) = \widehat I \;,
\label{e:green1}
\end{equation}
where $\widehat \Sigma_{1 , 2}(E)$
are the selfenergies introduced by the two semiinfinite leads \cite{Da95}.
The explicit form of the retarded selfenergy due to the mode of
wavevector $k_y$ is:
\begin{equation}
\Sigma(E)={{1 \over 2} \left( E -\epsilon(k_y)- i
\sqrt {4-(E-\epsilon(k_y))^2} \right) } \;,
\label{e:sigma}
\end{equation}
for energies within its band $|E -\epsilon(k_y)| < 2$, where
$\epsilon(k_y)=2{\rm cos}(k_y)$ is
the eigenenergy of the mode $k_y$ which is quantized as
$k_y=(n_{k_y}\pi)/(W+1)$,
$n_{k_y}$ being an integer from 1 to $W$. The transformation from the normal
modes
to the local tight--binding basis is obtained from the amplitudes of the
normal modes, $<n|k_y> = \sqrt{2/(W+1)}{\rm sin}(nk_y)$.
Note that in writing Eq.~(\ref{e:sigma}) we assumed that the magnetic field 
was zero outside the dot \cite{Da95}.

\subsection{Numerical Procedures}
Input/output leads were attached at opposite corners of the dot
as follows: input lead connected from site $(1,1)$ to 
site $(1,1+W)$, and output lead from  site $(L,1)$ to site $(L,1+W)$.
We have checked that changing the sites at which leads are attached does not
qualitatively modify the results discussed here. The conductance was averaged 
over disorder realizations (local distribution of vacancies) and within 
selected energy ranges. The latter were chosen to fit the
number of channels in the leads. More specifically, for
leads with $N_{\rm ch}$ channels  energy averages were carried  in the range,
\begin{equation}
E \in [E_{N_{\rm ch}},E_{N_{\rm ch}+1}]\;,
\end{equation}
\noindent for $N_{\rm ch}$ channels in the leads, where,
\begin{equation}
E_n=-2\left (1+{\rm cos}\frac{\pi n}{W+1}\right )\;.
\end{equation}
Some calculations were also carried out at a fixed Fermi energy. In all cases 
averages were done over at least 1200 values of the conductance.

\section{Results}
\subsection{Zero field conductance}
\begin{figure}
\begin{picture}(236,200) (-5,-15)
\epsfbox{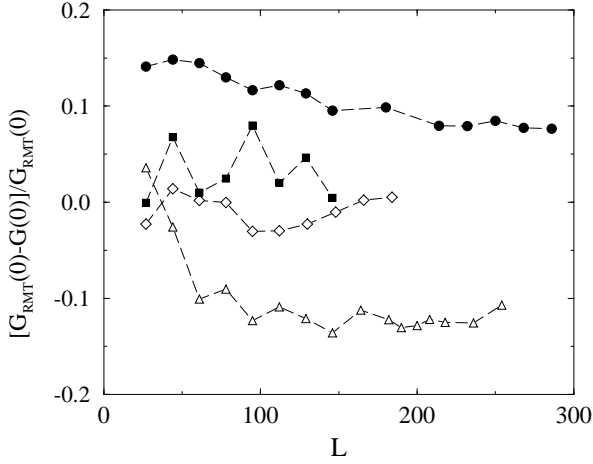}
\end{picture}
\caption{
Relative deviations of the numerical results for the zero field conductance 
with respect to the RMT result versus dot
size $L$.  Leads of width $W$ were attached at  opposite 
corners of the dot. Chaoticity was induced by 
introducing $L$ bulk vacancies (see text).
The results correspond to averages over 60 disorder realizations and 
21 energies in the ranges corresponding to the number
of channels $N_{\rm ch}$ in the leads (see text). The results correspond
to ($W$,$N_{\rm ch}$) = (1,1) -diamonds-, (3,3) -squares-, (18,9) 
-circles- and (9,9) -triangles-. The lines are guides to the eye.
\label{f:zero_w}}
\end{figure}

Fig.~\ref{f:zero_w} shows relative deviations of the conductance with 
respect to the RMT result (see Eq.~(\ref{e:RMTnch})) for narrow and rather 
wide leads as a function of 
the dot size $L$. It is noted that for small $W$ deviations are always smaller 
than 5\%, and typically below 2\%. 
The results fluctuate more appreciably for the narrower lead ($W$=1) 
as expected \cite{LV00}.
Relative deviations from the RMT result are significantly larger for 
$W=$ 9 and 18. The results of Fig.~\ref{f:zero_w} suggest that the
difference with respect RMT is not a size effect. The
larger deviation, and stronger variation in the explored range of $L$, 
observed for $N_{\rm ch}=W=9$ is likely
a consequence of the important contribution that the center of the band 
($E=0$) has
in that case. Both the center of the band and its bottom ($E=-4$) show
rather odd behaviors. In particular at $E=0$ no weak localization effect
was observed (see below).

\begin{figure}
\begin{picture}(236,200) (-5,-15)
\epsfbox{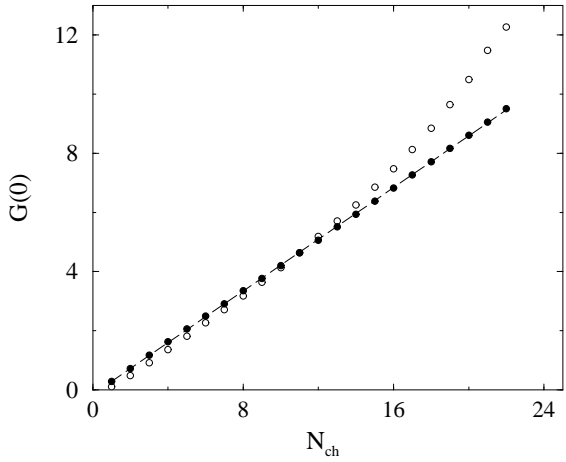}
\end{picture}
\caption{
Zero field conductance $G(0)$ in units of the conductance quantum
in dots of linear size $L$=95 versus 
the number of channels in the leads $N_{\rm ch}$.
Leads of width $W=$ 22 (empty circles) and $W=2N_{\rm ch}$ (filled circles)
were attached at  opposite 
corners of dots of linear size $L=78$. Chaoticity was induced by 
introducing $L$ bulk vacancies (see text).
The results correspond to averages over 60 disorder realizations and 
21 energies in the ranges corresponding to the number
of channels $N_{\rm ch}$ in the leads (see text). 
The straight line (broken curve) fitted to the results for $W=2N_{\rm ch}$ is:
$G(0)=0.44N_{\rm ch}-0.15$.
\label{f:zero_nch}}
\end{figure}

The change in the zero field conductance as the number of channels is
varied, for fixed dot size, is illustrated in 
Fig.~\ref{f:zero_nch}. The results for $N_{\rm ch}=W/2$ can be accurately
fitted by means of a straight line, as expected, although the slope
is smaller than the RMT prediction (see caption of Fig.~\ref{f:zero_nch}
and Eq.~\ref{e:RMTnch}).
The slope of the straight line varies with the ratio  $N_{\rm ch}/W$, 
or alternatively
the average Fermi energy; for instance at $E=0$ it is actually larger 
than 0.5. For fixed leads width $W=22$ and a variable number of 
channels a large deviation with respect to a straight line is 
instead observed. This deviation, which is a
consequence of the concomitant change in the Fermi energy as
the number of channels is varied,  increases
with the number of channels, likely due to the increasing importance of the
contribution of the band center. Numerical results indicate that
at the band center the conductance shows a much stronger dependence (increase)
on the dot size that at any other energy within the band, 
probably due to the building up
of the singularity in the density of sates characteristic of the square
lattice at that energy. These results suggest that if the $N_{\rm ch}$
dependence has to be investigated it is more reliable to work
at a fixed $N_{\rm ch}/W$ ratio and vary the leads width.

\subsection{Magnetoconductance: Weak Localization}
\begin{figure}
\begin{picture}(236,276) (30,-15)
\epsfbox{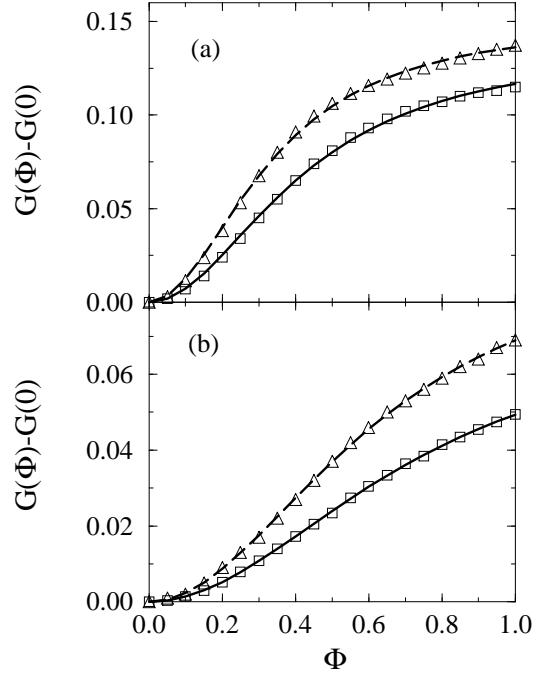}
\end{picture}
\caption{
Magnetoconductance  as a function of the
magnetic flux (in units of their respective quanta) in $78 \times 78$ chaotic 
cavities with leads of width $W$ = 22 (a) and 44 (b)
attached at  opposite corners of the dot, as discussed in the text. 
Chaoticity was induced by introducing $L=78$ bulk vacancies (see text).
Averages were taken over 60 disorder realizations
and  21 energies in the ranges corresponding to the number of channels 
$N_{\rm ch}$ in the leads. The numerical results correspond to 
$N_{\rm ch}$ = 4 (triangles) and 12 (squares) and  were fitted by means of 
Eq.~({\protect \ref{e:peak}}) with the parameters reported in Table I.
\label{f:conduc_energy}}
\end{figure}

We first discuss the energy dependence of the critical flux and
of the weak localization term. 
This was done by investigating rather large $W$ and  varying the
number of channels in each lead. This is equivalent to vary
the energy range over which the energy was calculated (see above).
Fig.~\ref{f:conduc_energy} depicts numerical results for cavities 
of linear size $L$=78 and leads of width $W$= 22 (a) and 44 (b).
The conductance was obtained by averaging over 60 disorder realizations and
21 energies in the ranges corresponding to the number of channels
in the leads (see subsection IIC).  The weak localization peak 
shows the expected behavior. The numerical results were fitted 
by means of Eq.~(\ref{e:peak}). At this stage it is worth noting that the 
conductance remains constant in a wide
range of fluxes only for small $W$. For large $W$ 
the conductance follows Eq. ~(\ref{e:peak}) in a rather narrow range of $\Phi$
and then increases steadily. This deviation from the Lorentzian--like law
hinders the fitting of the numerical results. The fitted parameters 
are reported in Table I. The parameters derived from
RMT (Eqs. (\ref{e:RMTab}--\ref{e:RMTb})) are also given in the Table.
Eq.~(\ref{e:RMTb}) was used with $k=1$ as  its explicit dependence on energy 
was not given in Ref. \cite{PW94}; this will suffice, however,
to illustrate our point concerning the strong energy dependence 
of that constant. We first note that the weak localization term is
smaller than the values predicted by RMT likely due to the large values
of $W$ (see below). However, the dependence of $a/b$ on the number of channels
is the correct one (it increases with $N_{\rm ch}$) but for
$W=44$ and $N_{\rm ch}=28$. The latter deviation is a consequence of the 
increasing importance of the band center. At that energy the results indicate
that $G(\Phi)$ decreases as a function of the flux, i.e., there is no
weak localization effect. Due to computing limitations
we have not been able to check whether this result is a size effect.
The numerical results for parameter $b$ indicate that it increases with
the  number of open channels, a behavior opposite to
that given by  RMT  with $k$= constant. This suggests that
it cannot be safely concluded that the critical flux is proportional to the
square root of the number of open channels, as, increasing the Fermi
energy, not only increases $N_{ch}$ but it also dramatically changes
constant $k$ in Eq.~(\ref{e:RMTb}). The energy dependence of $k$
already appears in the supersymmetric $\sigma$--model result \cite{Ef97}. 
We have checked that if the energy dependent factor in Eq.~(\ref{e:b}),
namely, $\nu D_0$, is included (the mean free path  was calculated
following the procedure of \cite{LC97,CL97}, see also \cite{Sh95})
the dependence of the RMT result on the number of channels is reversed, 
in agreement with our numerical results.

\begin{figure}
\begin{picture}(236,200) (-5,-15)
\epsfbox{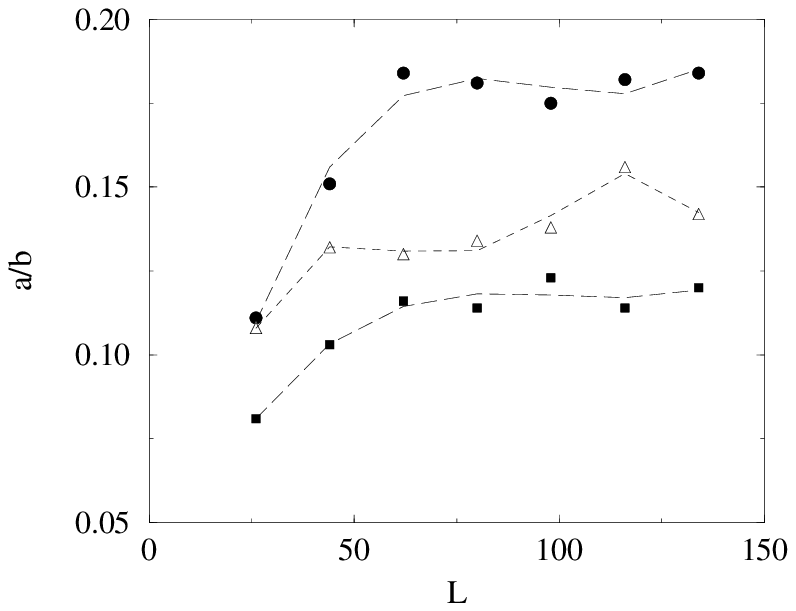}
\end{picture}
\caption{
Weak localization term ($a/b$ in Eq.~({\protect \ref{e:peak}})) as a
function of the dot size. Leads of width $W$ were attached at  opposite 
corners of the dot. Chaoticity was induced by 
introducing $L$ bulk vacancies (see text).
Averages were taken over 60 disorder realizations and 
21 energies in the ranges corresponding to the number
of channels $N_{\rm ch}$ in the leads (see text). The results correspond
to ($W$,$N_{\rm ch}$) = (10,5) -circles-, (2,1) -triangles- and (1,1) 
-squares-. The lines are guides to the eye.
\label{f:a:b_l}}
\end{figure}

\begin{figure}
\begin{picture}(236,200) (-5,-15)
\epsfbox{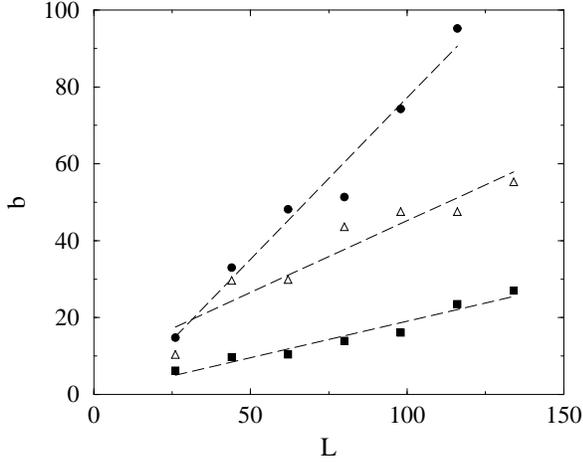}
\end{picture}
\caption{
Same as Fig.~(\ref{f:a:b_l}) for  constant $b$ in Eq.~(\ref{e:peak}).
The fitted straight lines are also shown.
\label{f:b_l}}
\end{figure}

The effects of the dot size on the weak localization peak 
were investigated for $L$ in the range of $L=27$--137 and three combinations
of $(W,N_{\rm ch})$. Averages were identical to those mentioned in the 
preceding paragraph. The results are depicted in Figs.~\ref{f:a:b_l} and
(\ref{f:b_l}). The weak localization term ($a/b$) shows a slight size
dependence at small $L$, saturating for $L$ approximately larger than 50
(see Fig.~\ref{f:a:b_l}). This indicates that the smaller 
values of $a/b$ obtained in our calculations, with respect to RMT, 
is probably not a size effect. The results for $(W,N_{\rm ch})$=(1,1) are 
slightly smaller than those for (2,1) surely due to the contribution
of the band center in the first case. The results for (10,5) are larger
than the other two, in agreement with RMT. On the other hand our results
for constant $b$ increase with $L$ as expected (see Eq.~(\ref{e:b})). 
The numerical results can 
be reasonably fitted by means of straight lines as shown in Fig.~\ref{f:b_l}.
The differences in the slopes is a consequence of the energy dependence  
discussed above. 

\begin{figure}
\begin{picture}(236,200) (-5,-15)
\epsfbox{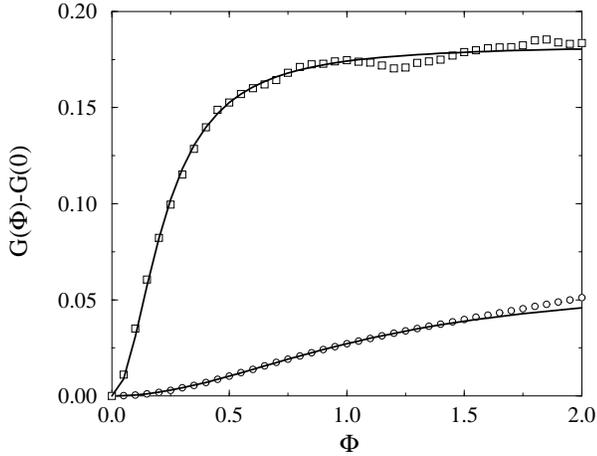}
\end{picture}
\caption{
Magnetoconductance  as a function of the
magnetic flux (in units of their respective quanta) in $78 \times 78$ chaotic 
cavities with leads of width $W$ = 4  and 78 (squares and circles, 
respectively) 
attached at  opposite corners of the dot, as discussed in the text. 
Chaoticity was induced by introducing $L=78$ bulk vacancies (see text).
The results correspond to averages over 1260 disorder realizations
and a fixed energy $E$=-2.001 which roughly correspond to 
$N_{\rm ch}=W/2$. The numerical results were
fitted by means of Eq.~({\protect \ref{e:peak}})
\label{f:conduc_nch}}
\end{figure}

\begin{figure}
\begin{picture}(236,200) (-5,-15)
\epsfbox{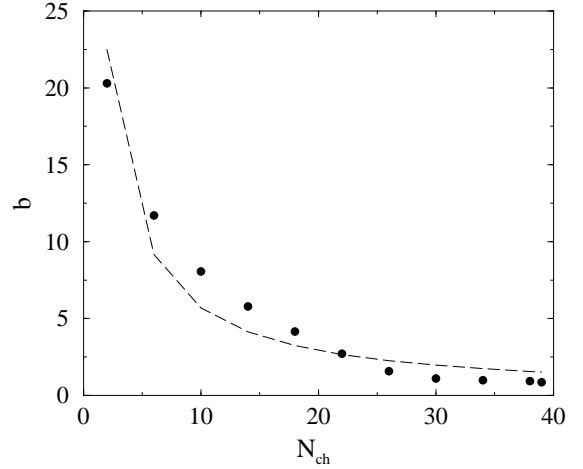}
\end{picture}
\caption{
Constant $b$ in Eq.~(\ref{e:peak})  as a
function of the number of channels in the leads $N_{\rm ch}$. Leads of width 
$W$ were attached at  opposite 
corners of dots of linear size $L$=78. Chaoticity was induced by 
introducing $L$ bulk vacancies (see text).
The results correspond to averages over 1260 disorder realizations,
$W$=4--78 and a fixed energy $E$=-2.001 (which roughly corresponds to
$N_{\rm ch}=W/2$). The broken line is the RMT result obtained from 
Eq.~({\protect \ref{e:RMTb}}) with $k=15$.
\label{f:b_nch}}
\end{figure}

\begin{figure}
\begin{picture}(236,200) (-5,-15)
\epsfbox{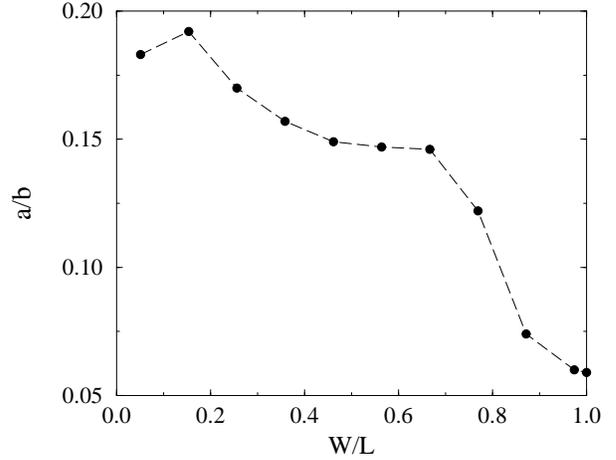}
\end{picture}
\caption{
Weak localization term ($a/b$ in Eq.~({\protect \ref{e:peak}})) as a
function of the leads width $W$. The leads  were attached at  opposite 
corners of dots of linear size $L$=78. Chaoticity was induced by 
introducing $L$ bulk vacancies (see text).
The results correspond to averages over 1260 disorder realizations,
$W$=4--78 and a fixed energy $E$=-2.001. The line is a guide to the eye.
\label{f:a:b_w:l}}
\end{figure}

In order to get rid as much as possible of the strong  energy 
dependence of the shape of the weak localization peak, we have carried out
the study of the effects of the leads width at a fixed energy. We have
chosen $E=-2.001$ (away from the band center and bottom) which
approximately correspond to $N_{\rm ch}=W/2$. We fixed the dot size at $L=78$
and varied the leads width in the range $W$=4--78. The results are
shown in Figs.~\ref{f:conduc_nch}-\ref{f:a:b_w:l}. The 
conductance versus the magnetic field for small and large $W$ is depicted
in Fig.(\ref{f:conduc_nch}). It is readily noted that both the
weak localization term and constant $b$ (or the inverse of the square
root of the critical flux) sharply decreases with $W$. Although
the results are nicely fitted by means of Eq.~\ref{e:peak}) the deviation
of the numerical results with respect to that equation which occurs at large 
$W$ (see above) is already observed for $W=78$ (note
that the fitting closely follows the numerical results only up to $\Phi 
\approx 1.5$). The decrease of $b$ with $W$, or, alternatively,
with $N_{\rm ch}$ is illustrated in Fig. \ref{f:b_nch}. The results can be 
satisfactorily fitted by means of the RMT result (see caption of
Fig. \ref{f:b_nch}). On the other hand the weak localization term shows
a size dependence that has not been previously anticipated. At small
$W$ (or number of channels) it increases as predicted by Eq.~(\ref{e:RMTab}).
However, beyond $W \approx 0.2L$ it begins to decrease sharply reaching a
value slightly larger than 0.05 for $W=L$.  To explore the possibility
that the weak localization term vanishes in the large $L$ limit we
have calculated the magnetoconductance for $W=L$, $E=-2.001$ and $L$
in the range 30--126. The numerical results were fitted by means
of Eq.~(\ref{e:peak}) with the parameters reported in Table II.
The results clearly indicate that $a/b$ does not vanish as $L$ increases.
The fact that $b$ is almost independent of $L$ is a consequence of
the dependence of $b$ on the ratio $L/N_{\rm ch}$ (note that by taking 
$W=L$ and a fixed energy the number of channels is proportional to $L$).

\section{Concluding Remarks}
Summarizing, we have presented what we believe to be the first detailed
numerical study of the effects of Fermi energy, leads width and dot size 
on the shape of the weak localization peak in quantum chaotic cavities.
The study was carried out on a model that was recently proposed
by us \cite{VL99} which shows all the expected features of closed 
chaotic quantum billiards.
Although the conclusions of our investigation qualitatively agree with 
most predictions of 
random matrix theory, some significant issues have to be highlighted.
We first note that our results show that  albeit the critical flux
is proportional to the square root of the number of open channels, 
as predicted by RMT, the
proportionality constant strongly depends on the Fermi energy in
agreement with Efetov's analysis \cite{Ef97}. This introduces a 
model (system) dependence which makes  theoretical (experimental)
comparisons with RMT rather delicate. Our results clearly illustrate the
size dependence of the critical flux, in particular 
$\Phi_c \propto 1/\sqrt{L}$, in agreement
with Efetov results \cite{Ef97} and the RMT results reported in
Refs. \cite{FP95,PW95} (note that this size dependence was not found
in a previously published RMT study, see Ref.\cite{PW94}). Finally, we 
have investigated the effects of the leads width concluding that the weak
localization term sharply decreases with the ratio $W/L$ (a
result that has not been previously reported), although
it is likely that it does not vanish in the infinite $L$ limit.
This suggest that RMT is probably not valid for sufficiently open systems.

\acknowledgments
This work was supported in part by the Spanish CICYT (grants PB96-0085
and 1FD97--1358).
Useful discussions with E. Cuevas and M. Ortu\~no are gratefully acknowledged.

\begin{table}
\caption{
Fittings of numerical results such as those of 
Fig.~{\protect \ref{f:conduc_energy}}
by means of Eq.~({\protect \ref{e:peak}}) for
$78 \times 78$ chaotic cavities with leads of widths $W$=22 and 44 
attached at  opposite corners of the dot, as discussed in the text. 
Chaoticity was induced by introducing $L$ vacancies within the bulk.
The magnetoconductance was obtained by averaging over 60 disorder 
realizations and 21 energies in the ranges corresponding to the number
of channels $N_{\rm ch}$ in the leads (see text).  
RMT results as obtained from Eqs.~({\protect \ref{e:RMTab}}) and 
({\protect \ref{e:RMTb}}) with $k=1$ are also reported.} 

\label{t:energy}
\begin{tabular}{llccccc}
 &  & \multicolumn{3}{c} {numerical} & \multicolumn{2}{c}{RMT} \\
\tableline
 $W$ & $N_{\rm ch}$ & $G(0)$ & $b$ & $a/b$ & $b$ & $a/b$ \\
\tableline
  22 &  4 & 1.32 & 5.62 & 0.137 & 0.88 & 0.22  \\ 
     & 12 & 5.12 & 8.96 & 0.151 & 0.32 & 0.24  \\ 
\tableline
  44  &  4 &  0.98 & 1.83 & 0.076 & 0.88 & 0.22  \\ 
      & 12 &  4.62 & 2.46 & 0.097 & 0.32 & 0.24  \\ 
      & 20 &  8.47 & 2.81 & 0.125 & 0.18 & 0.24  \\ 
      & 28 & 12.87 & 4.73 & 0.091 & 0.14 & 0.25  \\ 
\end{tabular}
\end{table}

\begin{table}
\caption{
Fittings of numerical results such as those of 
Fig.~{\protect \ref{f:conduc_energy}}
by means of Eq.~({\protect \ref{e:peak}}) for
$L \times L$ chaotic cavities with leads of widths $W=L$ 
attached at  opposite corners of the dot, as discussed in the text. 
Chaoticity was induced by introducing $L$ vacancies within the bulk.
The magnetoconductance was obtained by averaging over 1260 disorder 
and at a fixed energy $E$=-2.001 (which roughly corresponds to
$N_{\rm ch}=W/2$).} 
\label{t:size}
\begin{tabular}{lccc}
 $L$ &  $G(0)$ & $b$ & $a/b$  \\
\tableline
  30 &  6.50   & 0.66 & 0.058  \\ 
  42 &  9.34   & 0.89 & 0.051  \\ 
  54 & 12.19   & 0.91 & 0.056  \\ 
  66 & 15.04   & 0.76 & 0.070  \\ 
  78 & 17.91   & 0.85 & 0.059  \\ 
 102 & 23.65   & 0.98 & 0.053  \\ 
 114 & 26.49   & 0.87 & 0.054  \\ 
 126 & 29.37   & 0.88 & 0.047  \\ 
\end{tabular}
\end{table}

%

\end{document}